\documentclass[11pt]{article}
\usepackage{amssymb,latexsym,amsmath,amsbsy}
\usepackage[dvips]{graphicx}
\usepackage{xcolor}
\usepackage{cite}
\usepackage{stmaryrd}  
\usepackage{tikz}
\usetikzlibrary{arrows, decorations.markings}
\allowdisplaybreaks
\def\nn{\nonumber}

\def\qdots{\mathinner{\mkern1mu\raise1pt\vbox{\kern7pt\hbox{.}}\mkern2mu \raise4pt\hbox{.}\mkern2mu\raise7pt\hbox{.}\mkern1mu}}
\def\Z{{\mathbb Z}}

\def\C{{\mathbb C}}

\def\g{\mathfrak{g}}

\def\so{\mathfrak{so}}

\def\lb{\llbracket}
\def\rb{\rrbracket}

\renewcommand{\theequation}{\arabic{section}.{\arabic{equation}}}
\newtheorem{theo}{Theorem}

\newtheorem{prop}[theo]{Proposition}
\def\mybox{\hfill$\Box$}%
\begin{document}
\begin{center}
{\Large \bf
$\Z_2^3$-grading of the Lie algebra $G_2$ and related color algebras} \\[5mm]
{\bf N.I.~Stoilova}\footnote{E-mail: stoilova@inrne.bas.bg}\\[1mm] 
Institute for Nuclear Research and Nuclear Energy, Bulgarian Academy of Sciencies,\\ 
Boul.\ Tsarigradsko Chaussee 72, 1784 Sofia, Bulgaria\\[2mm] 
{\bf J.\ Van der Jeugt}\footnote{E-mail: Joris.VanderJeugt@UGent.be}\\[1mm]
Department of Mathematics, Computer Science and Statistics, Ghent University,\\
Krijgslaan 281-S9, B-9000 Gent, Belgium.
\end{center}

\vskip 2 cm

\begin{abstract}
\noindent 
We present a special and attractive basis for the exceptional Lie algebra $G_2$, which turns $G_2$ into a $\Z_2^3$-graded Lie algebra.
There are two basis elements for each degree of $\Z_2^3\setminus\{(0,0,0)\}$, thus yielding 14 basis elements.
We give a general and simple closed form expression for commutators between these basis elements.
Next, we use this $\Z_2^3$-grading in order to examine graded color algebras.
Our analysis yields three different $\Z_2^3$-graded color algebras of type $G_2$.
Since the $\Z_2^3$-grading is not compatible with a Cartan-Weyl basis of $G_2$,
we also study another grading of $G_2$.
This is a $\Z_2^2$-grading, compatible with a Cartan-Weyl basis, and for which we can
also construct a $\Z_2^2$-graded color algebra of type $G_2$.
\end{abstract}

\vskip 10mm
\noindent $\Z_2^3$-grading and coloring of $G_2$ 


\setcounter{equation}{0}
\section{Introduction} \label{sec:A}%

Color algebras and color superalgebras were introduced by Rittenberg and Wyler~\cite{Rit1,Rit2}.
Such algebras are graded by some abelian grading group $\Gamma$, and the simplest case not coinciding with a Lie algebra or Lie superalgebra is for $\Gamma=\Z_2\times\Z_2$.
For an algebra graded by $\Z_2\times\Z_2=\Z_2^2$, there are already two distinct choices for the Lie bracket~\cite{Rit1,Rit2,Scheunert}: 
these are now referred to as $\Z_2\times\Z_2$-graded Lie algebras and $\Z_2\times\Z_2$-graded Lie superalgebras
(this terminology - though common in literature - is slightly misleading, since these algebras are in general not Lie algebras nor Lie superalgebras).

Applications of $\Z_2\times\Z_2$-graded Lie (super)algebras in physics were rare for many years~\cite{LR1978,Vasiliev1985}.
But since the recognition of a $\Z_2\times\Z_2$-graded Lie superalgebra underlying the symmetries of L\'evy–Leblond equations~\cite{Aizawa1}, 
these $\Z_2\times\Z_2$-graded algebras have experienced a revival in mathematical physics.
They appeared in graded (quantum) mechanics and quantization,
in $\Z_2\times\Z_2$-graded two-dimensional models,
in $\Z_2\times\Z_2$-graded superspace formulations
and in particular in parastatistics and in the description and application of other types of parabosons and parafermions
(see e.g.\ \cite{Bruce2,AMD2020,Aizawa5,Quesne2021,Aizawa6} and references therein).

In~\cite{SV2023} we constructed classes of $\Z_2\times\Z_2$-graded (color) Lie algebras corresponding to classical Lie algebras of type $A_n$, $B_n$, $C_n$ and $D_n$.
In that construction, the defining matrix form of the basis elements of the graded color algebra is the same as the matrix form of the basis elements of the classical Lie algebra up to certain sign changes, and the basis of the Cartan subalgebra remains unchanged.
The technique of~\cite{SV2023} is rather tricky, and did not lead to any results for exceptional Lie algebras.
In~\cite{SV2024} we constructed $\Z_2\times\Z_2$-graded (color) Lie superalgebras corresponding to basic classical Lie superalgebras of type $A(m,n)$, $B(m,n)$, $C(n)$ and $D(m,n)$
using the same technique, but again we could not treat exceptional Lie superalgebras with this method.
On the other hand, exceptional Lie algebras like $G_2$ or exceptional Lie superalgebras like $G(3)$ - and their colorings - 
prove to be useful in superconformal quantum mechanics~\cite{Toppan-G3,AZK2018}.

Furthermore, colorings of Lie (superalgebras) based on a grading group of type $\Z_2^3$ or generally $\Z_2^n$ became of interest~\cite{DA2021,AAD2020,Balbino2024}.
The last paper, Ref.~\cite{Balbino2024}, gave us inspiration to reinvestigate possible colorings of the exceptional Lie algebra $G_2$.

The renewed activity in color Lie (super)algebras in the field of mathematical physics overlooked that in recent years there has also been a lot of activity in the study of graded Lie algebras by pure mathematicians. 
The state of the art was summarized in the book ``Gradings on simple Lie algebras''~\cite{EK13}, and a lot of research has followed up.
For the Lie algebra $G_2$, all gradings were determined in~\cite{DM06}, and we shall relate the two particular gradings appearing in the current paper with those of~\cite{DM06}.
Independently, the gradings of $G_2$ were found in~\cite{BT09}.
In the same context, gradings of the compact Lie algebra $G_2$ and $F_4$ were studied in~\cite{CDM10}, and of $G_2$ and $D_4$ in~\cite{EK18}.

In the current paper, we first investigate a grading of the ordinary Lie algebra $G_2$ by the grading group $\Gamma=\Z_2^3$ in Section~2.
Deleting the neutral element in $\Gamma$ gives rise to a grading set $\Gamma^*$ consisting of 7 elements.
It is well known that these 7 elements form a natural labeling for the 7 points of the Fano plane~\cite{Fano1,Fano2,Baez}.
We present a $\Z_2^3$-grading of $G_2$, with two basis elements for each of the 7 elements of $\Gamma^*$ (yielding 14 basis elements), 
or for each of the 7 points of the Fano plane.
Of course, there is a well established indirect connection between $G_2$ and the Fano plane: 
$G_2$ is the Lie algebra of derivations of octonions, for which the multiplication rules are also determined by the Fano plane.
In this paper, however, we give a direct definition of $G_2$ in terms of a novel basis $A_\alpha^\zeta$ ($\alpha,\zeta\in\Gamma^*$) 
with simple commutation relations~\eqref{AAA}, see Proposition~2.
The grading of $G_2$ by the grading group $\Gamma=\Z_2^3$ is not new, and can be found in~\cite[Theorem 2, case (25)]{DM06}.
A convenient basis for this grading was constructed in~\cite{DF24} and in~\cite{CD24}.
Although the basis $A_\alpha^\zeta$ ($\alpha,\zeta\in\Gamma^*$) can be brought in one-to-one correspondence with that of~\cite{CD24}, it has still the benefit of being nice and clean with uniformity in its commutation relations.

Our $\Z_2^3$-graded basis for $G_2$ allows us to investigate compatible colorings of this Lie algebra.
We find three different $\Z_2^3$-graded color Lie algebras of type $G_2$ in Section~3.
Each of these color Lie algebras has a basis consisting of two elements of degree $\alpha$, for every $\alpha\in\Gamma^*$.
But the graded bracket (consisting of commutators and anti-commutators) is different in each case.
For all these cases, we list the bracket relations among the 14 basis elements explicitly.
Clearly, this is somewhat tedious.
But we hope it helps the reader to verify the result, to understand the difference between the Lie algebra $G_2$ and a color Lie algebra of type $G_2$,
and to understand the difference between the three distinct colorings.
Moreover, such explicit basis elements and brackets can be used in applications such as (super)conformal quantum mechanics~\cite{Toppan-G3}
or integrable vertex models~\cite{Martins} based on exceptional Lie or color algebras.

The standard matrix representation of the basis elements for $G_2$ or the corresponding color algebras consists of anti-symmetric matrices.
For some purposes, it is more convenient to have a matrix representation in which the Cartan subalgebra elements are diagonal.
Such a representation is given in Section~4.
There, it is shown that this yields a $\Z_2^2$-grading of $G_2$, with two basis elements of degree $(0,0)$, four of degree $(1,0)$, 
four of degree $(0,1)$ and four of degree $(1,1)$.
 
This grading of $G_2$ can be found in~\cite[Theorem 2, case (23)]{DM06},
and can also be colored, yielding a $\Z_2^2$-graded color Lie algebra of type $G_2$
(or a $\Z_2^2$-graded Lie algebra of type $G_2$ in the common terminology).

\section{$\Z_2^3$-grading of the Lie algebras $\so(7)$ and $G_2$}
\setcounter{equation}{0} \label{sec:B}

\subsection{The grading group $\Gamma$ and the oriented Fano plane}

Let $\Gamma=\Z_2\times\Z_2\times\Z_2=\Z_2^3$, for which the elements are written as
\begin{equation}
\Gamma=\{ 000, 100, 010, 110, 001, 101, 011, 111\}
\end{equation}
and $\Gamma^*=\Gamma \setminus \{000\}$.
So we identify $\alpha=(\alpha_1,\alpha_2,\alpha_3)$ with the string $\alpha_1\alpha_2\alpha_3$.
For elements $\alpha$ and $\beta$ of $\Gamma$ or $\Gamma^*$, we denote addition in $\Z_2^3$ by $\alpha+\beta$,
for example $110+011=101$. We also define
\begin{equation}
(\cdot|\cdot) : \Gamma^*\times\Gamma^* \rightarrow \Z_2 : 
(\alpha|\beta)= \alpha_1\beta_1 + \alpha_2\beta_2 +\alpha_3\beta_3,
\end{equation}
where addition is in $\Z_2$; for example $(110|111) = 1+1+0=0$.

The set of all (complex) $8\times 8$-matrices is denoted by ${\sf M}_8(\C)\equiv {\sf M}$.
Rows and columns of these matrices are labeled by the elements of $\Gamma$, in the fixed order
\begin{equation}
000\quad 100\quad 010\quad 110\quad 001\quad 101\quad 011\quad 111.
\label{index}
\end{equation}
${\sf M}$ is a vector space and an algebra by matrix multiplication.
${\sf M}$ is also a $\Gamma$-graded algebra:
\begin{equation}
{\sf M}= \bigoplus_{\alpha\in \Gamma} {\sf M}_\alpha
\end{equation}
where a matrix $X\in{\sf M}$ with $X=(X_{\beta\gamma})$ ($\beta,\gamma\in\Gamma$) satisfies
\begin{equation}
X\in{\sf M}_\alpha \qquad\Leftrightarrow\qquad X_{\beta\gamma}=0 \hbox{ unless } \beta+\gamma=\alpha.
\end{equation}
If $X\in{\sf M}_\alpha$ then we say that the degree of $X$ is $\alpha$.
It is clear that ${\sf M}_\alpha \cdot {\sf M}_\beta \subset {\sf M}_{\alpha+\beta}$.
Note that the matrix form of the eight subspaces ${\sf M}_\alpha$ ($\alpha\in\Gamma$) coincides with those of Appendix~D in~\cite{Balbino2024}.
For example,
\begin{equation}
{\sf M}_{010}: \hbox{matrices of shape } 
\left(
\begin{array}{cccccccc}
 & &*& & & & & \\
 & & &*& & & & \\
*& & & & & & & \\
 &*& & & & & & \\
 & & & & & &*& \\
 & & & & & & &*\\
 & & & &*& & & \\
 & & & & &*& & 
\end{array} \right),
\end{equation}
where the symbol $*$ denotes which entries of the $8\times 8$-matrix can be nonvanishing.

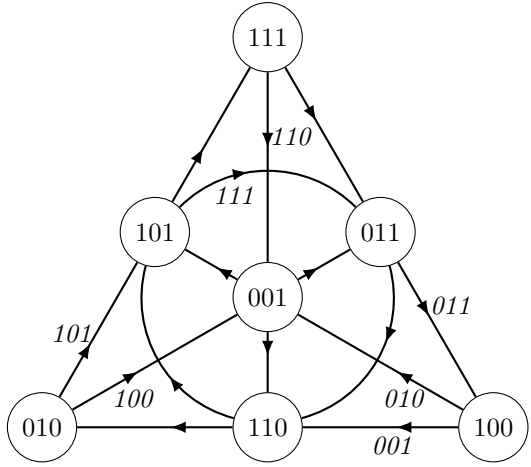
\begin{figure}[htb]
\begin{center}
\begin{tikzpicture}[scale = 1.5,decoration={markings,mark=at position 0.4 with {\arrow[line width = 1pt]{latex}}}]
\node[circle,draw] (1) at (0, 1.155) {\small $001$}; 
\node[circle,draw] (2) at  (0, 3.46) {\small $111$}; 
\node[circle,draw] (3) at (2, 0) {\small $100$};
\node[circle,draw] (4) at  (-2,0) {\small $010$};
\node[circle,draw] (5) at (0,0) {\small $110$};
\node[circle,draw] (6) at (1,1.732) {\small $011$};  
\node[circle,draw] (7) at (-1,1.732) {\small $101$};  
\draw (1.1,0) node[below] {\small \em 001};
\draw (1.2,0.42) node[below] {\small \em 010};
\draw (1.62,1.25) node[below] {\small \em 011};
\draw (-1.2,0.42) node[below] {\small \em 100};
\draw (-1.74,1) node[below] {\small \em 101};
\draw (-0.07,2.58) node[right] {\small \em 110};
\draw (-0.3,2.22) node[below] {\small \em 111};
\draw [thick, postaction={decorate}] (2)--(1);
\draw [thick, postaction={decorate}] (1)--(5);
\draw [thick, postaction={decorate}] (2)--(6);
\draw [thick, postaction={decorate}] (6)--(3);
\draw [thick, postaction={decorate}] (3)--(5);
\draw [thick, postaction={decorate}] (5)--(4);
\draw [thick, postaction={decorate}] (4)--(7);
\draw [thick, postaction={decorate}] (7)--(2);
\draw [thick, postaction={decorate}] (1)--(7);
\draw [thick, postaction={decorate}] (3)--(1);
\draw [thick, postaction={decorate}] (4)--(1);
\draw [thick, postaction={decorate}] (1)--(6);
\draw [thick, postaction={decorate}, bend left = 45] (7) to (6);
\draw [thick, postaction={decorate}, bend left = 45] (6) to (5);
\draw [thick, postaction={decorate}, bend left = 45] (5) to (7);
\end{tikzpicture}
\end{center}
\caption{The oriented Fano plane, with the 7 points labeled by encircled elements of $\Gamma^*$, 
and the 7 lines labeled in italic by the elements of $\Gamma^*$.}
\end{figure}

For the elements of $\Gamma^*$, it will be useful to represent them in an oriented Fano plane, as in Figure~1.
The seven points in this plane are labeled by an element of $\Gamma^*$~\cite{Fano1,Fano2,Baez}.
Three different points $\alpha$, $\beta$, $\gamma$ are on a line if and only if $\alpha+\beta+\gamma=000$.
Also the seven lines of the Fano plane are labeled by the elements of $\Gamma^*$.
A point $\alpha$ lies on a line $\zeta$ if and only if $(\alpha|\zeta)=0$.
For every point $\alpha$, there are exactly three different lines (labeled by $\lambda$, $\mu$, $\nu$) through $\alpha$.
These are the three elements $\lambda$, $\mu$, $\nu$ of $\Gamma^*$ such that 
$(\alpha|\lambda)=(\alpha|\mu)=(\alpha|\nu)=0$.
We shall denote this set of elements as
\begin{equation}
\alpha^\perp = \{ \lambda,\mu,\nu\} = \{ \zeta\in\Gamma^* | (\alpha|\zeta)=0 \}.
\end{equation}
Conversely, for every $\lambda\in\Gamma^*$, there are three different points $\alpha$, $\beta$, $\gamma$ on the line labeled by $\lambda$.
This line through $\alpha$, $\beta$ and $\gamma$ will sometimes be denoted by $L(\alpha,\beta,\gamma)$.
These points satisfy $(\lambda|\alpha) = (\lambda|\beta) = (\lambda|\gamma) = 0$.
For every two distinct points $\alpha$ and $\beta$, there is a unique line through $\alpha$ and $\beta$.
Let us denote the label of this line by $\ell(\alpha,\beta)\in\Gamma^*$.
Then
\begin{equation}
\mu=\ell(\alpha,\beta) \qquad\Leftrightarrow\qquad (\mu|\alpha) = (\mu|\beta) = (\mu|\alpha+\beta) = 0.
\end{equation}
Note that our notation is somewhat superfluous, since $\ell(\alpha,\beta)=L(\alpha,\beta,\alpha+\beta)$, but it will be useful to keep both notions.

Further on, the orientation of the lines of the Fano plane -- as indicated by arrows in Figure~1 -- will play a role.
If $\alpha$, $\beta$, $\gamma$ are three points on an oriented line (thus $\alpha+\beta+\gamma=000$), then
\begin{equation}
\begin{tabular}{ll}
$\sigma(\alpha,\beta,\gamma)=+1$ & if $(\alpha, \beta, \gamma)$ follows the orientation;\\
$\sigma(\alpha,\beta,\gamma)=-1$ & if $(\alpha, \beta, \gamma)$ is in the opposite direction.
\end{tabular}
\end{equation}
For example, $\sigma(110,010,100)=+1$, $\sigma(110,101,011)=+1$, $\sigma(111,100,011)=-1$.

Finally, note that the oriented Fano plane also fixes the product of the seven non-scalar basis elements $e_\alpha$ ($\alpha\in\Gamma^*$) of the octonion algebra~\cite{Baez} by
\begin{equation} 
e_\alpha \cdot e_\beta = \sigma(\alpha,\beta,\alpha+\beta) e_{\alpha+\beta}, \qquad \alpha, \beta \in\Gamma^*,\ \alpha\ne\beta,
\label{octonions}
\end{equation}
see also~\cite{E1998}.

In the following, this terminology of the oriented Fano plane will turn out to be useful.

\subsection{Basis of $\so(7)$ and $G_2$}

The $\Z_2^3$-graded matrices of ${\sf M}$ will be used to define a basis for the Lie algebras $\so(7)$ and its subalgebra $G_2$.
For convenience, we continue to work with $8\times 8$-matrices (instead of $7\times 7$-matrices), and $\so(7)$ consists of anti-symmetric matrices of ${\sf M}$ with
the first row and the first column zero.
In general, let $E_{\alpha\beta}$ be the $8\times 8$-matrix with a 1 at position $(\alpha,\beta)$ and zeroes elsewhere.
Let
\[
m_{\alpha\beta}=E_{\alpha\beta}-E_{\beta\alpha}\qquad (\alpha,\beta\in\Gamma),
\]
with $m_{\beta\alpha}=-m_{\alpha\beta}$.
The 21 matrices $m_{\alpha\beta}$ with $\alpha,\beta\in\Gamma^*$ and $\alpha<\beta$ (according to the order~\eqref{index}).
form a basis for $\so(7)$.
This is a classical basis, and the commutators are given by
\begin{equation}
[m_{\alpha\beta}, m_{\mu\nu}]=\delta_{\beta\mu}m_{\alpha\nu}- \delta_{\alpha\nu}m_{\mu\beta}- \delta_{\beta\nu}m_{\alpha\mu}+ \delta_{\alpha\mu}m_{\nu\beta}.
\label{mm}
\end{equation}
This yields a $\Z_2^3$-grading of $\so(7)$: 
\[
\so(7)=\bigoplus_{\alpha\in\Gamma} \so(7)_\alpha,
\]
where $\so(7)_{000}=\{0\}$ has dimension 0, and each $\so(7)_\alpha$ has dimension 3 for $\alpha\in\Gamma^*$ (giving total dimension 21 for $\so(7)$). 
A basis for $\so(7)_\alpha$ is given by 
\begin{equation}
m_{\beta\gamma}, m_{\beta'\gamma'}, m_{\beta''\gamma''}, 
\end{equation}
where $\beta+\gamma= \beta'+\gamma'= \beta''+\gamma''=\alpha$; otherwise said: 
$L(\alpha,\beta,\gamma)$, $L(\alpha,\beta',\gamma')$ and $L(\alpha,\beta'',\gamma'')$ are the three distinct lines through the point $\alpha$ in the Fano plane.

Next, we introduce a nice and symmetric basis for the exceptional Lie algebra $G_2$, as a subalgebra of $\so(7)$.
For every $\alpha\in\Gamma^*$, let us define three matrices
\begin{equation}
A_\alpha^\zeta,\qquad \zeta\in\alpha^\perp,
\end{equation}
i.e.\ for every line $\zeta$ through the point $\alpha$, we determine a matrix $A_\alpha^\zeta$ of degree $\alpha$.
When $\alpha^\perp=\{\lambda,\mu,\nu\}$, these three lines can be denoted as 
\begin{equation}
\lambda=L(\alpha,\beta,\gamma),\quad \mu=L(\alpha,\beta',\gamma'), \quad \nu=L(\alpha,\beta'',\gamma''),
\end{equation} 
where the order is chosen such that $\sigma(\alpha,\beta,\gamma)=\sigma(\alpha,\beta',\gamma')=\sigma(\alpha,\beta'',\gamma'')=+1$.
Then, under the assumption that $\sigma(\lambda,\mu,\nu)=+1$, 
\begin{equation}
A_\alpha^\lambda = m_{\beta'\gamma'} - m_{\beta''\gamma''}, \quad
A_\alpha^\mu = m_{\beta''\gamma''} - m_{\beta\gamma}, \quad
A_\alpha^\nu = m_{\beta\gamma} - m_{\beta'\gamma'}.
\label{defA}
\end{equation}
For example, the three matrices of degree $100$ are given by
\[
A_{100}^{001}=m_{111,011}-m_{001,101},\quad A_{100}^{011}=m_{001,101}-m_{110,010},\quad A_{100}^{010}=m_{110,010}-m_{111,011}.
\]
From the general definition, it is clear that
\begin{equation}
\sum_{\zeta\in\alpha^\perp} A_\alpha^\zeta = 0 \qquad\hbox{for every }\alpha\in\Gamma^*.
\label{Asum}
\end{equation}
Hence there are only 14 linearly independent elements in the set of 21 matrices $\{ A_\alpha^\zeta | \alpha\in\Gamma^*, \zeta\in\alpha^\perp\}$.
We shall show that this yields a basis for the 14-dimensional Lie algebra $G_2$.
The main ingredient for this result is the commutation relation among the 21 matrices $A_\alpha^\zeta$.
This is a particularly nice and original formula, with the complete list of $G_2$ commutators expressed in a symmetric and closed form expression.
It can be seen as a counterpart of the $G_2$ basis given in~\cite[Section~1.30]{Dictionary},
where the constraints are more involved (including a particular antisymmetric tensor that can be related to octonions),
and where the commutation relations also contain terms with this antisymmetric tensor.

It should be added that the matrices $\{ A_\alpha^\zeta | \alpha\in\Gamma^*, \zeta\in\alpha^\perp\}$ can be brought in one-to-one correspondence with the $G_2$ algebra realization  considered in~\cite{CD24}.
The elements $\varphi_{ij}$ of~\cite[Section 2.2]{CD24} correspond to our matrices $m_{\alpha\beta}$.
The condition $a_1+a_2+a_3=0$ in~\cite[Lemma 2.6]{CD24} selects linear combinations of the elements~\eqref{defA}.
The main merit of our presentation is the choice of our elements $A_\alpha^\zeta$ and the following proposition.

\begin{prop}
The 21 matrices $A_\alpha^\lambda$, with $\alpha\in\Gamma^*$ and $\lambda\in\alpha^\perp$ satisfy:
\begin{equation}
[A_\alpha^\lambda, A_\beta^\mu]=0 \quad\hbox{ when } \alpha=\beta,
\label{AA}
\end{equation}
otherwise (under the assumption that $\sigma(\alpha,\beta,\alpha+\beta)=+1$) 
\begin{equation}
[A_\alpha^\lambda, A_\beta^\mu]=
\begin{cases}
-2A_{\alpha+\beta}^\lambda \quad\hbox{ when } \lambda=\mu=\ell(\alpha,\beta),\\[2mm]
A_{\alpha+\beta}^{\ell(\alpha,\beta)} \quad\quad\hbox{ when } \lambda=\ell(\alpha,\beta)\ne\mu \hbox{ or } \lambda\ne\ell(\alpha,\beta)=\mu,\\[2mm]
A_{\alpha+\beta}^{\lambda+\mu} \quad\quad\hbox{ otherwise}.
\end{cases}
\label{AAA}
\end{equation}
\end{prop}

\noindent {\bf Proof.} 
When $\alpha=\beta$, the commutator $[A_\alpha^\lambda, A_\beta^\mu]$ should be of degree $\alpha+\alpha=000$ by the $\Z_2^3$-grading of the matrices.
Since $\so(7)_{000}=\{0\}$, this commutator vanishes.\\
Consider next the commutator $[A_\alpha^\lambda, A_\beta^\mu]$  with $\lambda=\mu=\ell(\alpha,\beta)$.
Let $\lambda=L(\alpha,\beta,\gamma)$ with orientation $\sigma(\alpha,\beta,\gamma)=+1$ (otherwise, one switches $A_\alpha^\lambda$ and $A_\beta^\mu$ in the commutator).
By definition
\[
A_\alpha^\lambda = m_{\beta'\gamma'} - m_{\beta''\gamma''},
\]
where $\alpha=\beta+\gamma=\beta'+\gamma'=\beta''+\gamma''$ and $\lambda=L(\alpha,\beta,\gamma)$, $\mu=L(\alpha,\beta',\gamma')$ and $\nu=L(\alpha,\beta'',\gamma'')$ are the three lines through the point $\alpha$, with $\sigma(\alpha,\beta,\gamma)=\sigma(\alpha,\beta',\gamma')=\sigma(\alpha,\beta'',\gamma'')=+1$.
Then the three lines through the point $\beta$ are given by $\lambda=L(\beta,\gamma,\alpha)$, $\mu'=L(\beta,\beta'',\beta')$ and $\nu'=L(\beta,\gamma',\gamma'')$,
where $\sigma(\beta,\gamma,\alpha)=\sigma(\beta,\beta'',\beta')=\sigma(\beta,\gamma',\gamma'')=+1$.
Therefore
\[
A_\beta^\lambda = m_{\beta''\beta'} - m_{\gamma'\gamma''}.
\]
Now it follows from~\eqref{mm} that
\begin{align*}
[A_\alpha^\lambda, A_\beta^\lambda] &= 
[ m_{\beta'\gamma'} - m_{\beta''\gamma''} , m_{\beta''\beta'} - m_{\gamma'\gamma''} ] \\
&= [m_{\beta'\gamma'}, m_{\beta''\beta'}]
- [m_{\beta'\gamma'}, m_{\gamma'\gamma''}] - [m_{\beta''\gamma''}, m_{\beta''\beta'}]
+ [m_{\beta''\gamma''}, m_{\gamma'\gamma''}] \\
&= -m_{\beta''\gamma'} -  m_{\beta'\gamma''}  -  m_{\beta'\gamma''} -  m_{\beta''\gamma'} = -2( m_{\beta'\gamma''}- m_{\gamma'\beta''}) .
\end{align*}
But the lines through the point $\gamma$ are $\lambda=L(\alpha,\beta,\gamma)$, $\mu''=L(\beta',\gamma'',\gamma)$ and $\nu''=L(\gamma', \beta'',\gamma)$,
with $\sigma(\alpha,\beta,\gamma)=\sigma(\beta',\gamma'',\gamma)=\sigma(\gamma', \beta'',\gamma)=+1$, and thus
\[
A_\gamma^\lambda =  m_{\beta'\gamma''} - m_{\gamma'\beta''},
\]
yielding $[A_\alpha^\lambda, A_\beta^\lambda] =-2 A_\gamma^\lambda$.

Using the same notation, the second case of~\eqref{AAA} to consider is $[A_\alpha^\lambda, A_\beta^{\mu'}]$ with $\mu'\ne\lambda$. One finds
\begin{align*}
[A_\alpha^\lambda, A_\beta^{\mu'}] 
&= [ m_{\beta'\gamma'} - m_{\beta''\gamma''} , m_{\gamma'\gamma''} - m_{\beta\gamma} ] \\
&=  m_{\beta'\gamma''}+m_{\beta''\gamma'}=  m_{\beta'\gamma''}-m_{\gamma'\beta''}=A_\gamma^\lambda.
\end{align*}

For the third commutator of~\eqref{AAA}, one finds:
\begin{align*}
[A_\alpha^\mu, A_\beta^{\mu'}] 
&= [ m_{\beta''\gamma''} - m_{\beta\gamma} , m_{\gamma'\gamma''} - m_{\gamma\alpha} ] \\
&=   -m_{\beta''\gamma'}+m_{\beta\alpha}=   m_{\gamma'\beta''}-m_{\alpha\beta}=A_\gamma^{\mu''},
\end{align*}
and since $\mu$, $\mu'$ and $\mu''$ are the three lines through the point $\beta'$, one has $\mu''=\mu+\mu'$.

\mybox

Now we have the following result.

\begin{prop}
Consider the 14-dimensional vector space $\g$ spanned by 21 elements $A_\alpha^\zeta$ ($\alpha\in\Gamma^*$, $\zeta\in\alpha^\perp$)
subject to the 7 linear relations~\eqref{Asum}.
Equipped with the commutation relations~\eqref{AA}-\eqref{AAA}, $\g$ is the Lie algebra $G_2$.
Note that $\g$ is $\Z_2^3$-graded:
\begin{equation}
\g=\sum_{\alpha\in\Gamma} \g_\alpha
\end{equation}
with $\g_{000}=\{0\}$ and $\dim\g_\alpha=2$ for each $\alpha\in\Gamma^*$.
\end{prop}

We shall sketch a proof of the above result in two ways.
The first way is probably less comfortable for mathematical physicist.
It uses the natural action of the matrices $A_\alpha^\zeta$ on the basis $e_\eta$ of octonions~\eqref{octonions}, where $\eta\in\Gamma$ and the order of the basis elements is determined by~\eqref{index}.
In the notation of~\eqref{defA}, this means 
\begin{equation}
A_\alpha^\lambda(e_{\beta'})=-e_{\gamma'}, \ A_\alpha^\lambda(e_{\gamma'})=e_{\beta'}, \ 
A_\alpha^\lambda(e_{\beta''})=e_{\gamma''}, \ A_\alpha^\lambda(e_{\gamma''})=-e_{\beta''}, 
\end{equation}
and $A_\alpha^\lambda(e_{\eta})=0$ for $\eta\in\Gamma\setminus\{\beta',\gamma', \beta'',\gamma''\}$.
Then one can check that the elements $A_\alpha^\zeta$ act as derivations on the algebra of octonions.
In fact, still following the notation of~\eqref{defA}, it is sufficient to check that $A_\alpha^\lambda$ acts as a derivation on the following 7 products:
\begin{align*}
&e_\alpha\cdot e_{\beta}=e_{\gamma},\quad e_\alpha\cdot e_{\beta'}=e_{\gamma'},
\quad e_\alpha\cdot e_{\beta''}=e_{\gamma''},\quad e_\beta\cdot e_{\beta'}=e_{\beta''},\\
&e_{\beta'}\cdot e_{\gamma}=e_{\gamma''},\quad e_{\gamma'}\cdot e_{\beta}=e_{\gamma''},\quad e_{\gamma'}\cdot e_{\gamma}=e_{\beta''};
\end{align*}
(the remaining ones involving $e_{000}$ being trivial).
The algebra of derivations of octonions is well known to be $G_2$~\cite{Baez}.

A more elaborate way -- but maybe more satisfying for mathematical physicists -- 
of identifying $\g$ with $G_2$ is to identify our basis $A_\alpha^\lambda$ with a known basis of $G_2$.
We shall do this explicitly, because this basis (and the commutation relations) will be used in the following section on colorings.
Let us choose 14 independent elements among the 21 elements $A_\alpha^\lambda$:
\begin{align}
&e_1=A_{100}^{010}, \ 
e_2=A_{100}^{001}, \ 
e_3=A_{010}^{100}, \ 
e_4=A_{010}^{001}, \ 
e_5=A_{110}^{110}, \ 
e_6=A_{110}^{001}, \ 
e_7=A_{001}^{100}, \nonumber \\
&e_8=A_{001}^{010}, \ 
e_9=A_{101}^{101},\ 
e_{10}=A_{101}^{010},\  
e_{11}=A_{011}^{100}, \ 
e_{12}=A_{011}^{011}, \ 
e_{13}=A_{111}^{101}, \ 
e_{14}=A_{111}^{011}.
\label{e_i}
\end{align}
So the order of the elements is chosen in such a way that $e_{2i-1}$ and $e_{2i}$ are of the same degree,
and the following lists a basis of each $\g_\alpha$ for $\alpha\in\Gamma$:
\begin{equation}
\begin{array}{llllllll}
000&100&010&110&001&101&011&111\\
\phantom{e_1, e_2\quad} & e_1, e_2\quad & e_3,e_4\quad & e_5,e_6\quad &e_7,e_8\quad  & e_9,e_{10}\quad  & e_{11},e_{12}\quad & e_{13},e_{14}
\end{array}
\label{gradinglist}
\end{equation}
In terms of the standard matrices, following~\eqref{defA}, these 14 elements read
\begin{align}
& e_1=E_{011,111}-E_{111,011}+E_{110,010}-E_{010,110}, \quad e_2=E_{101,001}-E_{001,101}+E_{111,011}-E_{011,111},\nonumber\\
& e_3=E_{100,110}-E_{110,100}+E_{111,101}-E_{101,111}, \quad e_4=E_{011,001}-E_{001,011}+E_{101,111}-E_{111,101},\nonumber\\
& e_5=E_{100,010}-E_{010,100}+E_{101,011}-E_{011,101}, \quad e_6=E_{111,001}-E_{001,111}+E_{011,101}-E_{101,011},\nonumber\\
& e_7=E_{100,101}-E_{101,100}+E_{110,111}-E_{111,110}, \quad e_8=E_{011,010}-E_{010,011}+E_{111,110}-E_{110,111},\nonumber\\
& e_9=E_{001,100}-E_{100,001}+E_{011,110}-E_{110,011}, \quad e_{10}=E_{111,010}-E_{010,111}+E_{110,011}-E_{011,110},\nonumber\\ 
& e_{11}=E_{111,100}-E_{100,111}+E_{110,101}-E_{101,110}, \quad e_{12}=E_{010,001}-E_{001,010}+E_{101,110}-E_{110,101},\nonumber\\
& e_{13}=E_{011,100}-E_{100,011}+E_{110,001}-E_{001,110},\quad e_{14}=E_{101,010}-E_{010,101}+E_{001,110}-E_{110,001}. 
\label{e_iE}
\end{align}
Then, following~\eqref{AA}-\eqref{AAA} and~\eqref{Asum} (or the matrix form above), the complete list of commutators among those 14 elements is easily computed:
\begin{align}
&[e_1 ,e_2]=0,\ [e_1 ,e_3]=-e_5 , \ [e_1 ,e_4]=-e_6 , \ [e_1 ,e_5]=e_3 , \ [e_1 ,e_6]=e_4 ,\  [e_1 ,e_7]=e_{10} , \nonumber\\
&[e_1 ,e_8]=-2e_{10} , \ [e_1 ,e_9]=-e_8 , \ [e_1 ,e_{10}]=2e_8 , \ [e_1 ,e_{11}]=e_{13}+e_{14} , \ [e_1 ,e_{12}]=-e_{14} , \nonumber\\ 
&[e_1 ,e_{13}]=-e_{11}-e_{12} , \ [e_1 ,e_{14}]=e_{12}, \ [e_2 ,e_3]=-e_6 , \ [e_2 ,e_4]=2e_6 , \ [e_2 ,e_5]=e_4 , \ [e_2 ,e_6]=-2e_4 ,\nonumber\\
&[e_2 ,e_7]=e_9 , \ [e_2 ,e_8]=e_{10} , \ [e_2 ,e_9]=-e_7 , \ [e_2 ,e_{10}]=-e_8 , \ [e_2 ,e_{11}]=-e_{13} , \ [e_2 ,e_{12}]=-e_{14} , \nonumber\\
&[e_2 ,e_{13}]=e_{11} , \ [e_2 ,e_{14}]=e_{12} ,\ [e_3 ,e_4]=0, \ [e_3 ,e_5]=-e_1 ,\ [e_3 ,e_6]=-e_2 ,\ [e_3 ,e_7]=-2e_{11} , \nonumber\\
&[e_3 ,e_8]=e_{11} , \ [e_3 ,e_9]=e_{13} , \ [e_3 ,e_{10}]=-e_{13}-e_{14} , \ [e_3 ,e_{11}]=2e_7 ,\ [e_3 ,e_{12}]=-e_7 , \nonumber\\
&[e_3 ,e_{13}]=-e_9 , \ [e_3 ,e_{14}]=e_{10}+e_9,\ [e_4 ,e_5]=-e_2 , \ [e_4 ,e_6]=2e_2 ,\ [e_4 ,e_7]=e_{11} ,\ [e_4 ,e_8]=e_{12} , \nonumber\\
&[e_4 ,e_9]=e_{13} , \ [e_4 ,e_{10}]=e_{14} , \ [e_4 ,e_{11}]=-e_7 , \ [e_4 ,e_{12}]=-e_8 , \ [e_4 ,e_{13}]=-e_9 , \ [e_4 ,e_{14}]=-e_{10} \nonumber\\
&[e_5 ,e_6]=0,\ [e_5 ,e_7]=e_{13} +e_{14} ,\ [e_5 ,e_8]=e_{13} +e_{14} , \ [e_5 ,e_9]=e_{12} ,\ [e_5 ,e_{10}]=e_{11} , \nonumber\\
&[e_5 ,e_{11}]=-e_{10} ,\ [e_5 ,e_{12}]=-e_9 ,\ [e_5 ,e_{13}]=-e_8 -e_7 ,\ [e_5 ,e_{14}]=-e_8 -e_7 , \ [e_6 ,e_7]=-e_{13} , \nonumber\\
&[e_6 ,e_8]=-e_{14} ,\ [e_6 ,e_9]=e_{11} ,\ [e_6 ,e_{10}]=e_{12} ,\ [e_6 ,e_{11}]=-e_9 ,\ [e_6 ,e_{12}]=-e_{10} ,\ [e_6 ,e_{13}]=e_7 , \nonumber\\
&[e_6 ,e_{14}]=e_8 ,\ [e_7 ,e_8]=0,\ [e_7 ,e_9]=e_2 , \ [e_7 ,e_{10}]=e_1 , \ [e_7 ,e_{11}]=-2e_3 , \ [e_7 ,e_{12}]=e_3 , \nonumber\\
&[e_7 ,e_{13}]=-e_6 , \ [e_7 ,e_{14}]=e_5 +e_6 ,\ [e_8 ,e_9]=e_1 , \ [e_8 ,e_{10}]=-2e_1 ,\ [e_8 ,e_{11}]=e_3 ,\ [e_8 ,e_{12}]=e_4 , \nonumber\\
&[e_8 ,e_{13}]=e_5 +e_6 , \ [e_8 ,e_{14}]=-e_6 ,\ [e_9 ,e_{10}]=0 , \ [e_9 ,e_{11}]=e_6 , \ [e_9 ,e_{12}]=e_5 , \nonumber\\
&[e_9 ,e_{13}]=2e_4 +2e_3 , \ [e_9 ,e_{14}]=-e_4 -e_3 , \ [e_{10} ,e_{11}]=e_5 , \ [e_{10} ,e_{12}]=e_6 ,\ [e_{10} ,e_{13}]=-e_4 -e_3 , \nonumber\\
&[e_{10} ,e_{14}]=e_4 , \ [e_{11} ,e_{12}]=0, \  [e_{11} ,e_{13}]=-e_2 , \ [e_{11} ,e_{14}]=e_2 +e_1 , \ [e_{12} ,e_{13}]=e_2 +e_1 , \nonumber\\
&[e_{12} ,e_{14}]=-2e_2 -2e_1 ,\ [e_{13} ,e_{14}]=0.
\label{comm-e_i}
\end{align}
One can now take as basis of a Cartan subalgebra of $\g$:
\begin{equation}
h_1=i e_2 -i e_1,\quad h_2=i e_1
\label{g2-h}
\end{equation}
and consider the following twelve root vectors with respect to this Cartan subalgebra:
\begin{align}
& x_1= e_3+\frac12 e_4+i e_5+\frac{i}{2}e_6, \quad
x_2= e_{11}+\frac12 e_{12}+i e_{13}+\frac{i}{2}e_{14}, \quad
x_3= e_7+\frac12 e_8+i e_9+\frac{i}{2}e_{10},\nn\\
& y_1= e_3+\frac12 e_4-i e_5-\frac{i}{2}e_6, \quad
y_2= e_{11}+\frac12 e_{12}-i e_{13}-\frac{i}{2}e_{14}, \quad
y_3= e_7+\frac12 e_8-i e_9-\frac{i}{2}e_{10},\nn\\
& a_{12}= \frac12 e_8-\frac{i}{2} e_{10}, \quad a_{13}=-\frac12 e_{12}+\frac{i}{2} e_{14}, \quad a_{23}=\frac12 e_4+\frac{i}{2} e_6,\nn\\
& a_{21}= -\frac12 e_8-\frac{i}{2} e_{10}, \quad a_{31}=\frac12 e_{12}+\frac{i}{2} e_{14}, \quad a_{32}=-\frac12 e_4+\frac{i}{2} e_6.
\label{g2-x}
\end{align}
The commutator table of these 14 elements is given in the Appendix.
Clearly, it is a Chevalley basis for $G_2$, with $y_2$ and $a_{12}$ as the two simple root vectors.
The table in the Appendix coincides with Table~1 of~\cite{g2-JLT}, up to the changes $h_1\rightarrow h_2$ and $h_2\rightarrow h_1+h_2$ in~\cite{g2-JLT}.
Note that the root vectors are not homogeneous for the $\Z_2^3$-grading, as they mix elements of distinct degree. 
In fact, it was already proved in~\cite{DM06} that the $\Z_2^3$-grading of $G_2$ is non-toral.

\setcounter{equation}{0}
\section{$\Z_2^3$-graded color Lie algebras of type $G_2$} 
\label{sec:C}%

For a general abelian additive grading group $\Gamma$, a mapping 
\[
\langle \cdot, \cdot\rangle : \Gamma\times\Gamma \rightarrow \Z_2 : (\alpha,\beta) \mapsto \langle\alpha,\beta\rangle
\]
is a sign factor~\cite{Balbino2024,Kobayashi1983} if for all elements $\alpha,\beta,\gamma$ of $\Gamma$
\[
\langle\beta,\alpha\rangle = \langle\alpha,\beta\rangle, \quad\hbox{ and }\quad 
\langle\alpha,\beta+\gamma\rangle =\langle\alpha,\beta\rangle + \langle\alpha,\gamma\rangle.
\]
Herein, equalities are in $\Z_2=\{0,1\}$, thus modulo 2.
Then the mapping $\epsilon : \Gamma\times\Gamma\rightarrow \{\pm 1\} : (\alpha,\beta) \mapsto (-1)^{\langle\alpha,\beta\rangle}$ 
is a commutation factor in the terminology of~\cite{Scheunert}, and is seen to be a symmetric bicharacter.

A $\Gamma$-graded color algebra $\g$ is now defined as follows~\cite{Rit1,Rit2,Scheunert,Balbino2024}.
Let $\displaystyle\g=\oplus_{\alpha\in\Gamma} \g_\alpha$ be a $\Gamma$-graded algebra with multiplication denoted by $\lb\cdot,\cdot\rb$,
hence $\lb \g_\alpha,\g_\beta\rb \subset \g_{\alpha+\beta}$. 
Then $\g$ is a $\Gamma$-graded color algebra if the symmetry and Jacobi identity are satisfied:
\begin{align}
& \lb x_{\alpha}, y_{\beta} \rb = -(-1)^{\langle\alpha,\beta\rangle} 
\lb y_{\beta}, x_{\alpha} \rb, \label{symmetry}\\
& \lb x_{\alpha}, \lb y_{\beta}, z_{\gamma}\rb \rb =
\lb \lb x_{\alpha}, y_{\beta}\rb , z_{\gamma} \rb +
(-1)^{\langle\alpha,\beta\rangle} \lb y_{\beta}, \lb x_{\alpha}, z_{\gamma}\rb \rb,
\label{jacobi}
\end{align} 
for all $x_\alpha\in\g_\alpha$, $y_\beta\in\g_\beta$, $z_\gamma\in\g_\gamma$.
As usual, an element $x\in\g_\alpha$ is called a homogeneous element of degree $\alpha$.

Let $\g$ be an associative $\Gamma$-graded algebra with a product denoted by $x\cdot y$, 
then the following bracket turns $\g$ into a $\Gamma$-graded color algebra:
\begin{equation}
\lb x_{\alpha} , y_{\beta}\rb = x_{\alpha} \cdot y_{\beta}- 
(-1)^{\langle\alpha,\beta\rangle} y_{\beta} \cdot x_{\alpha}\ ,
\label{ZZbracket}
\end{equation}
i.e.\ for a bracket derived from an associative product the Jacobi identity~\eqref{jacobi} is automatically satisfied.
Thus a bracket between homogeneous elements is always a commutator or an anti-commutator.

A $\Gamma$-graded color algebra such that $\langle\alpha,\alpha\rangle=0$ for all $\alpha\in\Gamma$ (but not all $\langle\alpha,\beta\rangle$ zero for $\alpha\ne\beta$) 
is often called a $\Gamma$-graded Lie algebra (following the ``misleading'' terminology mentioned in the Introduction).
If, on the other hand, there is at least one $\alpha$ with $\langle\alpha,\alpha\rangle=1$, then $\g$ is often called
a $\Gamma$-graded Lie superalgebra~\cite{Balbino2024}.
The difference between these notions is of course relevant when these algebras are interpreted as parastatistics algebras~\cite{Toppan2,Toppan3,Balbino2024}.

For $\Gamma=\Z_2^n$, a list of inequivalent sign factors has been given in~\cite[Appendix A]{Balbino2024}.
This is based on the classification of commutation factors for $\Gamma=\Z_2^n$ in~\cite[(5.20)-(5.21)]{Scheunert},
where equivalence of commutation factors for an abelian group $\Gamma$ is fixed in~\cite[Definition~4]{Scheunert}.
Since we are looking for colorings of $G_2$, for which we have already a novel $\Z_2^3$-grading in Section~2, 
we are particularly interested in the case $n=3$.
In the rest of this section, $\Gamma=\Z_2^3$ and $\Gamma^*=\Gamma\setminus\{000\}$.
For $\Z_2^3$, there are five inequivalent types of sign factors, denoted in~\cite[Appendix A]{Balbino2024}
by $3_1, 3_2, 3_3, 3_4$ and $3_5$. 
In the case of $3_1$, $\langle\alpha,\beta\rangle=0$ for all $\alpha,\beta\in\Gamma$, hence the $\Gamma$-graded color algebra is just an ordinary Lie algebra.
The $\Gamma$-graded ordinary Lie algebra $G_2$ has been determined in the previous section, in particular in Proposition~2.

Next, we investigated possible colorings of this $\Gamma$-graded Lie algebra $G_2$, in the following way.
Let $e_i$ ($i=1,\ldots,N$) be a $\Gamma$-graded basis of a $\Gamma$-graded Lie algebra $\g$ with structure constants determined by
\[
[e_i,e_j]= \sum_k c_{ij}^k e_k.
\]
For a fixed sign factor $\langle \cdot, \cdot\rangle$ on $\Gamma$, 
$\tilde\g$ is a $\Gamma$-graded color algebra on $\Gamma$ 
compatible with $\g$ if $\tilde\g$ has a $\Gamma$-graded basis $\tilde e_i$ ($i=1,\ldots,N$) 
and there is a bracket $\lb\cdot,\cdot\rb$ satisfying \eqref{symmetry}-\eqref{jacobi} with
\begin{equation}
\lb \tilde e_i,\tilde e_j \rb = \sum_k \epsilon_{ijk} c_{ij}^k \tilde e_k, \qquad\hbox{with } \epsilon_{ijk}\in\{+1,-1\}.
\label{compatible}
\end{equation}

For the case of $G_2$, we examined possible colorings for each possible sign factor, and found three different examples.
To be more precise, the colorings we analyzed are of the following type: 
for each $e_i$ of\eqref{e_iE}, we introduced all possible sign changes for the four nonzero elements of $e_i$ as a candidate matrix form of $\tilde e_i$.
The three examples thus found are all with sign factors of type $3_2$.
We found none with sign factors of type $3_3$, $3_4$ or $3_5$.
This is not surprising, as type $3_2$ corresponds to a $\Gamma$-graded color Lie algebra, whereas $3_3$, $3_4$ and $3_5$ would correspond to $\Gamma$-graded color Lie superalgebras.
Note also that for sign factors of type $3_3$, $3_4$ or $3_5$, there would be at least one element $\tilde e_i$ with $\lb \tilde e_i,\tilde e_i \rb = \{ \tilde e_i,\tilde e_i \} =0$,
and this cannot hold by making sign changes in $e_i$.

The restriction of our approach should be emphasized.
We do not obtain a classification of $\Gamma$-graded color algebras of type $G_2$ for each possible sign factor,
as we only examined those with sign changes in the $e_i$'s leading to brackets of type~\eqref{compatible}.
Despite this, the existence of such examples is worth knowing.

Let us present the three examples here.

\subsection{Case 1}

The first case is with a sign factor
\begin{equation}
\langle \alpha,\beta \rangle = \alpha_1\beta_2+\alpha_2\beta_1.
\label{sign12}
\end{equation}
Instead of denoting the new basis elements with $\tilde e_i$, let us (without causing any confusion) still denote them by $e_i$.
A matrix form of a basis $e_i$ ($i=1,\ldots,14$) of $\g$ in terms of the graded matrices of ${\sf M}$ is given by
\begin{align}
& e_1=E_{011,111}-E_{111,011}+E_{110,010}-E_{010,110}, \quad e_2=E_{101,001}-E_{001,101}+E_{111,011}-E_{011,111},\nonumber\\
& e_3=E_{100,110}-E_{110,100}+E_{111,101}-E_{101,111}, \quad e_4=E_{001,011}-E_{011,001}+E_{101,111}-E_{111,101},\nonumber\\
& e_5=E_{010,100}+E_{100,010}+E_{101,011}+E_{011,101}, \quad e_6=E_{111,001}+E_{001,111}-E_{101,011}-E_{011,101},\nonumber\\
& e_7=E_{100,101}-E_{101,100}+E_{110,111}-E_{111,110}, \quad e_8=E_{011,010}-E_{010,011}+E_{111,110}-E_{110,111},\nonumber\\
& e_9=E_{001,100}-E_{100,001}+E_{011,110}-E_{110,011}, \quad e_{10}=E_{111,010}-E_{010,111}+E_{110,011}-E_{011,110},\nonumber\\ 
& e_{11}=E_{111,100}-E_{100,111}+E_{110,101}-E_{101,110}, \quad e_{12}=E_{001,010}-E_{010,001}+E_{101,110}-E_{110,101},\nonumber\\
& e_{13}=-E_{011,100}-E_{100,011}+E_{001,110}+E_{110,001},\quad e_{14}=E_{010,101}+E_{101,010}-E_{001,110}-E_{110,001}. 
\label{e_i12}
\end{align}
Note the few sign changes compared to~\eqref{e_iE}, and that the grading is still according~\eqref{gradinglist}. 
Since all these basis elements are homogeneous, every bracket $\lb e_i,e_j\rb$ is either a commutator or an anti-commutator.
The complete set of brackets between these basis elements is given by
\begin{align}
&[e_1 ,e_2]=0,\ \{e_1 ,e_3\}=e_5 , \ \{e_1 ,e_4\}=e_6 , \ \{e_1 ,e_5\}=-e_3 , \ \{e_1 ,e_6\}=-e_4 ,\  [e_1 ,e_7]=e_{10} , \nonumber\\
&[e_1 ,e_8]=-2e_{10} , \ [e_1 ,e_9]=-e_8 , \ [e_1 ,e_{10}]=2e_8 , \ \{e_1 ,e_{11}\}=-e_{13}-e_{14} , \ \{e_1 ,e_{12}\}=e_{14} , \nonumber\\ 
&\{e_1 ,e_{13}\}=e_{11}+e_{12} , \ \{e_1 ,e_{14}\}=-e_{12}, \ \{e_2 ,e_3\}=e_6 , \ \{e_2 ,e_4\}=-2e_6 , \ \{e_2 ,e_5\}=-e_4 , 
\ \{e_2 ,e_6\}=2e_4 ,\nonumber\\
&[e_2 ,e_7]=e_9 , \ [e_2 ,e_8]=e_{10} , \ [e_2 ,e_9]=-e_7 , \ [e_2 ,e_{10}]=-e_8 , \ \{e_2 ,e_{11}\}=e_{13} , 
\ \{e_2 ,e_{12}\}=e_{14} , \nonumber\\
&\{e_2 ,e_{13}\}=-e_{11} , \ \{e_2 ,e_{14}\}=-e_{12} ,\ [e_3 ,e_4]=0, \ \{e_3 ,e_5\}=-e_1 ,\ \{e_3 ,e_6\}=-e_2 ,\ [e_3 ,e_7]=-2e_{11} , \nonumber\\
&[e_3 ,e_8]=e_{11} , \ \{e_3 ,e_9\}=e_{13} , \ \{e_3 ,e_{10}\}=-e_{13}-e_{14} , \ [e_3 ,e_{11}]=2e_7 ,\ [e_3 ,e_{12}]=-e_7 , \nonumber\\
&\{e_3 ,e_{13}\}=-e_9 , \ \{e_3 ,e_{14}\}=e_{10}+e_9,\ \{e_4 ,e_5\}=-e_2 , \ \{e_4 ,e_6\}=2e_2 ,\ [e_4 ,e_7]=e_{11} ,\ [e_4 ,e_8]=e_{12} , \nonumber\\
&\{e_4 ,e_9\}=e_{13} , \ \{e_4 ,e_{10}\}=e_{14} , \ [e_4 ,e_{11}]=-e_7 , \ [e_4 ,e_{12}]=-e_8 , \ \{e_4 ,e_{13}\}=-e_9 , 
\ \{e_4 ,e_{14}\}=-e_{10} \nonumber\\
&[e_5 ,e_6]=0,\ [e_5 ,e_7]=e_{13} +e_{14} ,\ [e_5 ,e_8]=e_{13} +e_{14} , \ \{e_5 ,e_9\}=e_{12} ,\ \{e_5 ,e_{10}\}=e_{11} , \nonumber\\
&\{e_5 ,e_{11}\}=e_{10} ,\ \{e_5 ,e_{12}\}=e_9 ,\ [e_5 ,e_{13}]=e_8 +e_7 ,\ [e_5 ,e_{14}]=e_8 +e_7 , \ [e_6 ,e_7]=-e_{13} , \nonumber\\
&[e_6 ,e_8]=-e_{14} ,\ \{e_6 ,e_9\}=e_{11} ,\ \{e_6 ,e_{10}\}=e_{12} ,\ \{e_6 ,e_{11}\}=e_9 ,\ \{e_6 ,e_{12}\}=e_{10} ,
\ [e_6 ,e_{13}]=-e_7 , \nonumber\\
&[e_6 ,e_{14}]=-e_8 ,\ [e_7 ,e_8]=0,\ [e_7 ,e_9]=e_2 , \ [e_7 ,e_{10}]=e_1 , \ [e_7 ,e_{11}]=-2e_3 , \ [e_7 ,e_{12}]=e_3 , \nonumber\\
&[e_7 ,e_{13}]=-e_6 , \ [e_7 ,e_{14}]=e_5 +e_6 ,\ [e_8 ,e_9]=e_1 , \ [e_8 ,e_{10}]=-2e_1 ,\ [e_8 ,e_{11}]=e_3 ,\ [e_8 ,e_{12}]=e_4 , \nonumber\\
&[e_8 ,e_{13}]=e_5 +e_6 , \ [e_8 ,e_{14}]=-e_6 ,\ [e_9 ,e_{10}]=0 , \ \{e_9 ,e_{11}\}=-e_6 , \ \{e_9 ,e_{12}\}=-e_5 , \nonumber\\
&\{e_9 ,e_{13}\}=-2e_4 -2e_3 , \ \{e_9 ,e_{14}\}=e_4 +e_3 , \ \{e_{10} ,e_{11}\}=-e_5 , \ \{e_{10} ,e_{12}\}=-e_6 ,
\ \{e_{10} ,e_{13}\}=e_4 +e_3 , \nonumber\\
&\{e_{10} ,e_{14}\}=-e_4 , \ [e_{11} ,e_{12}]=0, \  \{e_{11} ,e_{13}\}=-e_2 , \ \{e_{11} ,e_{14}\}=e_2 +e_1 , 
\ \{e_{12} ,e_{13}\}=e_2 +e_1 , \nonumber\\
&\{e_{12} ,e_{14}\}=-2e_2 -2e_1 ,\ [e_{13} ,e_{14}]=0.
\label{comm-e_i12}
\end{align}

We have checked these brackets using a simple computer algebra package.
But in fact, it is not too hard to work through them by hand using the explicit form~\eqref{e_i12}.
Thus, there is a $\Gamma$-graded color Lie algebra compatible with the $\Gamma$-graded ordinary Lie algebra $G_2$, 
with graded basis given by $\{ e_i | i=1,2,\ldots,14\}$, brackets given by~\eqref{comm-e_i12} and sign factor given by~\eqref{sign12}.
Note that~\eqref{symmetry} is satisfied, and that the Jacobi identity~\eqref{jacobi} is automatically satisfied since the brackets in~\eqref{comm-e_i12}
are satisfied by the matrices of~\eqref{e_i12}.

\subsection{Case 2}

The second case is with a sign factor
\begin{equation}
\langle \alpha,\beta \rangle = \alpha_1\beta_3+\alpha_3\beta_1.
\label{sign13}
\end{equation}
A matrix form of a basis of $\g$, which is still denoted by $e_i$ ($i=1,\ldots,14$), in terms of the graded matrices of ${\sf M}$ is given by
\begin{align}
& e_1=E_{011,111}-E_{111,011}+E_{110,010}-E_{010,110}, \quad e_2=E_{101,001}-E_{001,101}+E_{111,011}-E_{011,111},\nonumber\\
& e_3=E_{100,110}-E_{110,100}+E_{111,101}-E_{101,111}, \quad e_4=E_{011,001}-E_{001,011}+E_{101,111}-E_{111,101},\nonumber\\
& e_5=E_{100,010}-E_{010,100}+E_{101,011}-E_{011,101}, \quad e_6=E_{111,001}-E_{001,111}+E_{011,101}-E_{101,011},\nonumber\\
& e_7=E_{100,101}+E_{101,100}+E_{110,111}+E_{111,110}, \quad e_8=E_{011,010}+E_{010,011}-E_{111,110}-E_{110,111},\nonumber\\
& e_9=E_{001,100}-E_{100,001}+E_{011,110}-E_{110,011}, \quad e_{10}=E_{111,010}-E_{010,111}+E_{011,110}-E_{110,011},\nonumber\\ 
& e_{11}=E_{111,100}+E_{100,111}-E_{110,101}-E_{101,110}, \quad e_{12}=E_{001,010}+E_{010,001}+E_{101,110}+E_{110,101},\nonumber\\
& e_{13}=E_{011,100}-E_{100,011}+E_{110,001}-E_{001,110},\quad e_{14}=E_{010,101}-E_{101,010}+E_{001,110}-E_{110,001}. 
\label{e_i13}
\end{align}
By our approach, there are only certain sign changes compared to~\eqref{e_iE}, and the degree of these elements is the same.
Since all basis elements are homogeneous, the brackets are commutators or anti-commutators:
\begin{align}
&[e_1 ,e_2]=0,\ [e_1 ,e_3]=-e_5 , \ [e_1 ,e_4]=-e_6 , \ [e_1 ,e_5]=e_3 , \ [e_1 ,e_6]=e_4 ,\  \{e_1 ,e_7\}=e_{10} , \nonumber\\
&\{e_1 ,e_8\}=-2e_{10} , \ \{e_1 ,e_9\}=e_8 , \ \{e_1 ,e_{10}\}=2e_8 , \ \{e_1 ,e_{11}\}=e_{13}+e_{14} , 
\ \{e_1 ,e_{12}\}=-e_{14} , \nonumber\\ 
&\{e_1 ,e_{13}\}=-e_{11}-e_{12} , \ \{e_1 ,e_{14}\}=e_{12}, \ [e_2 ,e_3]=-e_6 , \ [e_2 ,e_4]=2e_6 , \ [e_2 ,e_5]=e_4 , 
\ [e_2 ,e_6]=-2e_4 ,\nonumber\\
&\{e_2 ,e_7\}=-e_9 , \ \{e_2 ,e_8\}=e_{10} , \ \{e_2 ,e_9\}=e_7 , \ \{e_2 ,e_{10}\}=-e_8 , \ \{e_2 ,e_{11}\}=-e_{13} , 
\ \{e_2 ,e_{12}\}=-e_{14} , \nonumber\\
&\{e_2 ,e_{13}\}=e_{11} , \ \{e_2 ,e_{14}\}=e_{12} ,\ [e_3 ,e_4]=0, \ [e_3 ,e_5]=-e_1 ,\ [e_3 ,e_6]=-e_2 ,\ [e_3 ,e_7]=2e_{11} , \nonumber\\
&[e_3 ,e_8]=-e_{11} , \ [e_3 ,e_9]=e_{13} , \ [e_3 ,e_{10}]=e_{13}+e_{14} , \ [e_3 ,e_{11}]=-2e_7 ,\ [e_3 ,e_{12}]=e_7 , \nonumber\\
&[e_3 ,e_{13}]=-e_9 , \ [e_3 ,e_{14}]=-e_{10}+e_9,\ [e_4 ,e_5]=-e_2 , \ [e_4 ,e_6]=2e_2 ,\ [e_4 ,e_7]=-e_{11} ,
\ [e_4 ,e_8]=-e_{12} , \nonumber\\
&[e_4 ,e_9]=e_{13} , \ [e_4 ,e_{10}]=-e_{14} , \ [e_4 ,e_{11}]=e_7 , \ [e_4 ,e_{12}]=e_8 , \ [e_4 ,e_{13}]=-e_9 , \ [e_4 ,e_{14}]=e_{10} \nonumber\\
&[e_5 ,e_6]=0,\ \{e_5 ,e_7\}=-e_{13} -e_{14} ,\ \{e_5 ,e_8\}=-e_{13} -e_{14} , \ \{e_5 ,e_9\}=e_{12} ,\ \{e_5 ,e_{10}\}=-e_{11} , \nonumber\\
&\{e_5 ,e_{11}\}=e_{10} ,\ \{e_5 ,e_{12}\}=-e_9 ,\ \{e_5 ,e_{13}\}=e_8 +e_7 ,\ \{e_5 ,e_{14}\}=e_8 +e_7 , 
\ \{e_6 ,e_7\}=e_{13} , \nonumber\\
&\{e_6 ,e_8\}=e_{14} ,\ \{e_6 ,e_9\}=e_{11} ,\ \{e_6 ,e_{10}\}=-e_{12} ,\ \{e_6 ,e_{11}\}=-e_9 ,\ \{e_6 ,e_{12}\}=e_{10} ,
\ \{e_6 ,e_{13}\}=-e_7 , \nonumber\\
&\{e_6 ,e_{14}\}=-e_8 ,\ [e_7 ,e_8]=0,\ \{e_7 ,e_9\}=-e_2 , \ \{e_7 ,e_{10}\}=e_1 , \ [e_7 ,e_{11}]=-2e_3 , 
\ [e_7 ,e_{12}]=e_3 , \nonumber\\
&\{e_7 ,e_{13}\}=e_6 , \ \{e_7 ,e_{14}\}=-e_5 -e_6 ,\ \{e_8 ,e_9\}=-e_1 , \ \{e_8 ,e_{10}\}=-2e_1 ,\ [e_8 ,e_{11}]=e_3 ,
\ [e_8 ,e_{12}]=e_4 , \nonumber\\
&\{e_8 ,e_{13}\}=-e_5 -e_6 , \ \{e_8 ,e_{14}\}=e_6 ,\ [e_9 ,e_{10}]=0 , \ \{e_9 ,e_{11}\}=-e_6 , \ \{e_9 ,e_{12}\}=-e_5 , \nonumber\\
&[e_9 ,e_{13}]=2e_4 +2e_3 , \ [e_9 ,e_{14}]=-e_4 -e_3 , \ \{e_{10} ,e_{11}\}=e_5 , \ \{e_{10} ,e_{12}\}=e_6 ,
\ [e_{10} ,e_{13}]=e_4 +e_3 , \nonumber\\
&[e_{10} ,e_{14}]=-e_4 , \ [e_{11} ,e_{12}]=0, \  \{e_{11} ,e_{13}\}=-e_2 , \ \{e_{11} ,e_{14}\}=e_2 +e_1 , 
\ \{e_{12} ,e_{13}\}=e_2 +e_1 , \nonumber\\
&\{e_{12} ,e_{14}\}=-2e_2 -2e_1 ,\ [e_{13} ,e_{14}]=0.
\label{comm-e_i13}
\end{align}
Thus this yields a $\Z_2^3$-graded color Lie algebra of type $G_2$, with sign factor~\eqref{sign13} of type $3_2$.

Note that the two commutation factors for $\Gamma$ following from~\eqref{sign12} and~\eqref{sign13} are equivalent in the sense of~\cite[Definition~4]{Scheunert}.
But the equivalence of the sign factors does not imply that the corresponding color algebras are isomorphic.
We have not been able to establish a color algebra isomorphism between~\eqref{comm-e_i12} and~\eqref{comm-e_i13}.

\subsection{Case 3}

The third case is with a sign factor
\begin{equation}
\langle \alpha,\beta \rangle = \alpha_2\beta_3+\alpha_3\beta_2.
\label{sign23}
\end{equation}
A matrix form of a basis $e_i$ ($i=1,\ldots,14$) of $\g$ in terms of the graded matrices of ${\sf M}$ is now given by
\begin{align}
& e_1=E_{011,111}-E_{111,011}+E_{110,010}-E_{010,110}, \quad e_2=E_{101,001}-E_{001,101}+E_{111,011}-E_{011,111},\nonumber\\
& e_3=E_{100,110}-E_{110,100}+E_{111,101}-E_{101,111}, \quad e_4=E_{011,001}-E_{001,011}+E_{101,111}-E_{111,101},\nonumber\\
& e_5=E_{100,010}-E_{010,100}+E_{101,011}-E_{011,101}, \quad e_6=E_{111,001}-E_{001,111}+E_{011,101}-E_{101,011},\nonumber\\
& e_7=E_{101,100}-E_{100,101}+E_{110,111}-E_{111,110}, \quad e_8=E_{011,010}-E_{010,011}+E_{111,110}-E_{110,111},\nonumber\\
& e_9=E_{100,001}-E_{001,100}+E_{011,110}-E_{110,011}, \quad e_{10}=E_{111,010}-E_{010,111}+E_{110,011}-E_{011,110},\nonumber\\ 
& e_{11}=E_{111,100}+E_{100,111}+E_{110,101}+E_{101,110}, \quad e_{12}=E_{001,010}+E_{010,001}-E_{101,110}-E_{110,101},\nonumber\\
& e_{13}=E_{011,100}+E_{100,011}+E_{110,001}+E_{001,110},\quad e_{14}=E_{010,101}+E_{101,010}+E_{001,110}+E_{110,001}. 
\label{e_i23}
\end{align}
The basis is again homogeneous, and like before there are only a few sign changes compared to~\eqref{e_iE}.
The complete list of brackets is given by:
\begin{align}
&[e_1 ,e_2]=0,\ [e_1 ,e_3]=-e_5 , \ [e_1 ,e_4]=-e_6 , \ [e_1 ,e_5]=e_3 , \ [e_1 ,e_6]=e_4 ,\  [e_1 ,e_7]=e_{10} , \nonumber\\
&[e_1 ,e_8]=-2e_{10} , \ [e_1 ,e_9]=-e_8 , \ [e_1 ,e_{10}]=2e_8 , \ [e_1 ,e_{11}]=e_{13}-e_{14} , \ [e_1 ,e_{12}]=e_{14} , \nonumber\\ 
&[e_1 ,e_{13}]=-e_{11}-e_{12} , \ [e_1 ,e_{14}]=-e_{12}, \ [e_2 ,e_3]=-e_6 , \ [e_2 ,e_4]=2e_6 , \ [e_2 ,e_5]=e_4 , \ [e_2 ,e_6]=-2e_4 ,\nonumber\\
&[e_2 ,e_7]=e_9 , \ [e_2 ,e_8]=e_{10} , \ [e_2 ,e_9]=-e_7 , \ [e_2 ,e_{10}]=-e_8 , \ [e_2 ,e_{11}]=-e_{13} , 
\ [e_2 ,e_{12}]=e_{14} , \nonumber\\
&[e_2 ,e_{13}]=e_{11} , \ [e_2 ,e_{14}]=-e_{12} ,\ [e_3 ,e_4]=0, \ [e_3 ,e_5]=-e_1 ,\ [e_3 ,e_6]=-e_2 ,
\ \{e_3 ,e_7\}=2e_{11} , \nonumber\\
&\{e_3 ,e_8\}=-e_{11} , \ \{e_3 ,e_9\}=-e_{13} , \ \{e_3 ,e_{10}\}=e_{13}-e_{14} , \ \{e_3 ,e_{11}\}=-2e_7 ,
\ \{e_3 ,e_{12}\}=e_7 , \nonumber\\
&\{e_3 ,e_{13}\}=e_9 , \ \{e_3 ,e_{14}\}=e_{10}+e_9,\ [e_4 ,e_5]=-e_2 , \ [e_4 ,e_6]=2e_2 ,\ \{e_4 ,e_7\}=-e_{11} ,
\ \{e_4 ,e_8\}=-e_{12} , \nonumber\\
&\{e_4 ,e_9\}=-e_{13} , \ \{e_4 ,e_{10}\}=e_{14} , \ \{e_4 ,e_{11}\}=e_7 , \ \{e_4 ,e_{12}\}=e_8 , 
\ \{e_4 ,e_{13}\}=e_9 , \ \{e_4 ,e_{14}\}=-e_{10} \nonumber\\
&[e_5 ,e_6]=0,\ \{e_5 ,e_7\}=-e_{13} +e_{14} ,\ \{e_5 ,e_8\}=-e_{13} +e_{14} , \ \{e_5 ,e_9\}=-e_{12} ,\ \{e_5 ,e_{10}\}=-e_{11} , \nonumber\\
&\{e_5 ,e_{11}\}=e_{10} ,\ \{e_5 ,e_{12}\}=e_9 ,\ \{e_5 ,e_{13}\}=e_8 +e_7 ,\ \{e_5 ,e_{14}\}=-e_8 -e_7 , 
\ \{e_6 ,e_7\}=e_{13} , \nonumber\\
&\{e_6 ,e_8\}=-e_{14} ,\ \{e_6 ,e_9\}=-e_{11} ,\ \{e_6 ,e_{10}\}=-e_{12} ,\ \{e_6 ,e_{11}\}=e_9 ,\ \{e_6 ,e_{12}\}=e_{10} ,
\ \{e_6 ,e_{13}\}=-e_7 , \nonumber\\
&\{e_6 ,e_{14}\}=e_8 ,\ [e_7 ,e_8]=0,\ [e_7 ,e_9]=e_2 , \ [e_7 ,e_{10}]=e_1 , \ \{e_7 ,e_{11}\}=-2e_3 , \ 
\{e_7 ,e_{12}\}=e_3 , \nonumber\\
&\{e_7 ,e_{13}\}=-e_6 , \ \{e_7 ,e_{14}\}=-e_5 -e_6 ,\ [e_8 ,e_9]=e_1 , \ [e_8 ,e_{10}]=-2e_1 ,\ \{e_8 ,e_{11}\}=e_3 ,
\ \{e_8 ,e_{12}\}=e_4 , \nonumber\\
&\{e_8 ,e_{13}\}=e_5 +e_6 , \ \{e_8 ,e_{14}\}=e_6 ,\ [e_9 ,e_{10}]=0 , \ \{e_9 ,e_{11}\}=e_6 , \ \{e_9 ,e_{12}\}=e_5 , \nonumber\\
&\{e_9 ,e_{13}\}=2e_4 +2e_3 , \ \{e_9 ,e_{14}\}=e_4 +e_3 , \ \{e_{10} ,e_{11}\}=e_5 , \ \{e_{10} ,e_{12}\}=e_6 ,
\ \{e_{10} ,e_{13}\}=-e_4 -e_3 , \nonumber\\
&\{e_{10} ,e_{14}\}=-e_4 , \ [e_{11} ,e_{12}]=0, \  [e_{11} ,e_{13}]=e_2 , \ [e_{11} ,e_{14}]=e_2 +e_1 , 
\ [e_{12} ,e_{13}]=-e_2 -e_1 , \nonumber\\
&[e_{12} ,e_{14}]=-2e_2 -2e_1 ,\ [e_{13} ,e_{14}]=0.
\label{comm-e_i23}
\end{align}
This case gives rise to a $\Z_2^3$-graded color Lie algebra of type $G_2$, with sign factor~\eqref{sign23} of type $3_2$.

\setcounter{equation}{0}
\section{$\Z_2^2$-graded color Lie algebra of type $G_2$ in the Cartan-Weyl basis} 
\label{sec:D}%

The $\Z_2^3$-gradings of the previous section were all in terms of $8\times 8$-matrices,
and the homogeneous basis elements were in general not corresponding to root vectors of $G_2$ (or to one of its colorings).
In this section, we will present a $\Z_2^2$-graded color Lie algebra of type $G_2$, 
but in such a way that the basis elements of the Cartan subalgebra correspond to diagonal matrices,
and the remaining homogeneous basis elements correspond to root vectors.

Note that the existence of a toral $\Z_2^2$-grading of $G_2$ was already established in~\cite[Theorem 2, case (23)]{DM06},
but no graded basis was given.

For this purpose, we first give such a basis for the ordinary Lie algebra $G_2$ in terms of $7\times 7$-matrices,
where the indices of the matrices are now just $1,2,\ldots,7$,
and $E_{ij}$ is the $7\times 7$-matrix with a 1 at position $(i,j)$ and zeroes elsewhere.

\begin{align}
& h_1=-E_{11}+2E_{22}-E_{33}+E_{44}-2E_{55}+E_{66},\quad h_2=E_{11}-E_{22}-E_{44}+E_{55},\nonumber \\
& x_1=E_{35}-E_{26}+\sqrt{2}E_{71}-\sqrt{2}E_{47}, \quad 
x_2=E_{16}-E_{34}+\sqrt{2}E_{72}-\sqrt{2}E_{57}, \nonumber\\
& x_3=-E_{15}+E_{24}+\sqrt{2}E_{73}-\sqrt{2}E_{67}, \quad 
y_1=-E_{53}+E_{62}-\sqrt{2}E_{17}+\sqrt{2}E_{74}, \nonumber\\
& y_2=-E_{61}+E_{43}-\sqrt{2}E_{27}+\sqrt{2}E_{75}, \quad
y_3=E_{51}-E_{42}-\sqrt{2}E_{37}+\sqrt{2}E_{76}, \nonumber\\
& a_{12}=E_{12}-E_{54},\quad a_{23}=E_{23}-E_{65},\quad a_{13}=E_{13}-E_{64},\nonumber\\
& a_{21}=E_{21}-E_{45},\quad a_{32}=E_{32}-E_{56},\quad a_{31}=E_{31}-E_{46}.
\label{xy-basis}
\end{align}
These matrices satisfy the commutation relations of the table given in the Appendix.
Note that this is a $\Z_2^2$-graded basis of $G_2$, with the degree of the elements given as follows:
\begin{equation}
\begin{array}{llll}
(0,0) & (0,1) & (1,0) & (1,1) \\
h_1, h_2\qquad\quad & x_1, y_1, a_{23}, a_{32}\quad & x_2, y_2, a_{13}, a_{31}\quad & x_3, y_3, a_{12}, a_{21}
\end{array}
\label{Z2Z2-G2}
\end{equation}

Also for this matrix representation, we examined the existence of a $\Z_2^2$-graded color algebra compatible with $G_2$.
We found a solution for this, again characterized by certain sign changes in the matrix elements of~\eqref{xy-basis}.
This yields a homogeneous basis for a $\Z_2^2$-graded color Lie algebra of type $G_2$:
\begin{align}
& h_1=-E_{11}+2E_{22}-E_{33}+E_{44}-2E_{55}+E_{66},\quad h_2=E_{11}-E_{22}-E_{44}+E_{55},\nonumber \\
& x_1=E_{35}-E_{26}+\sqrt{2}E_{71}-\sqrt{2}E_{47}, \quad 
x_2=-E_{16}-E_{34}+\sqrt{2}E_{72}+\sqrt{2}E_{57}, \nonumber\\
& x_3=E_{15}-E_{24}+\sqrt{2}E_{73}-\sqrt{2}E_{67}, \quad 
y_1=-E_{53}+E_{62}-\sqrt{2}E_{17}+\sqrt{2}E_{74}, \nonumber\\
& y_2=E_{61}+E_{43}-\sqrt{2}E_{27}-\sqrt{2}E_{75}, \quad
y_3=E_{51}-E_{42}+\sqrt{2}E_{37}-\sqrt{2}E_{76}, \nonumber\\
& a_{12}=E_{12}-E_{54},\quad a_{23}=E_{23}-E_{65},\quad a_{13}=E_{13}+E_{64},\nonumber\\
& a_{21}=E_{21}-E_{45},\quad a_{32}=E_{32}-E_{56},\quad a_{31}=E_{31}+E_{46}.
\label{xy-gradedbasis}
\end{align}
Since the basis is homogeneous, each bracket is a commutator or an anti-commutator. 
The complete list of brackets is given by:
\begin{align}
&[h_1,h_2]=0,\quad,[h_1,a_{12}]=-3a_{12},\quad [h_1,a_{13}]=0,\quad [h_1,a_{23}]=3a_{23},\quad [h_1,a_{21}]=3a_{21},\nn\\ 
&[h_1,a_{31}]=0,\quad [h_1,a_{32}]=-3a_{32},\quad [h_1,x_1]=x_1,\quad [h_1,x_2]=-2x_2,\quad [h_1,x_3]=x_3,\nn\\
&[h_1,y_1]=-y_1,\quad [h_1,y_2]=2y_2,\quad [h_1,y_3]=-y_3, \quad[h_2,a_{12}]=2a_{12},\quad [h_2,a_{13}]=a_{13},\nn\\
&[h_2,a_{23}]=-a_{23},\quad [h_2,a_{21}]=-2a_{21},\quad [h_2,a_{31}]=-a_{31},\quad [h_2,a_{32}]=a_{32},\quad [h_2,x_1]=-x_1,\nn\\
&[h_2,x_2]=x_2,\quad [h_2,x_3]=0,\quad [h_2,y_1]=y_1,\quad [h_2,y_2]=-y_2,\quad [h_2,y_3]=0\nn\\
&\{a_{12},a_{13}\}=0,\quad \{a_{12},a_{23}\}=a_{13},\quad [a_{12},a_{21}]=h_2,\quad \{a_{12},a_{31}\}=a_{32},\quad \{a_{12},a_{32}\}=0,\nn\\
&\{a_{12},x_1\}=x_2,\quad \{a_{12},x_2\}=0,\quad [a_{12},x_3]=0,\quad \{a_{12},y_1\}=0,\quad \{a_{12},y_2\}=y_1,\nn\\
&[a_{12},y_3]=0,\quad \{a_{13},a_{23}\}=0,\quad \{a_{13},a_{21}\}=a_{23},\quad [a_{13},a_{31}]=h_1+2h_2,\quad \{a_{13},a_{32}\}=a_{12},\nn\\ 
&\{a_{13},x_1\}=x_3,\quad [a_{13},x_2]=0,\quad \{a_{13},x_3\}=0,\quad \{a_{13},y_1\}=0,\quad [a_{13},y_2]=0,\nn\\
&\{a_{13},y_3\}=-y_1,\quad \{a_{23},a_{21}\}=0,\quad \{a_{23},a_{31}\}=a_{21},\quad [a_{23},a_{32}]=h_1+h_2,\quad [a_{23},x_1]=0,\nn\\
&\{a_{23},x_2\}=x_3,\quad \{a_{23},x_3\}=0,\quad [a_{23},y_1]=0,\quad \{a_{23},y_2\}=0,\quad \{a_{23},y_3\}=-y_2,\nn\\
&\{a_{21},a_{31}\}=0,\quad \{a_{21},a_{32}\}=a_{31},\quad \{a_{21},x_1\}=0,\quad \{a_{21},x_2\}=x_1,\quad [a_{21},x_3]=0,\nn\\ 
&\{a_{21},y_1\}=y_2,\quad \{a_{21},y_2\}=0,\quad [a_{21},y_3]=0,\quad \{a_{31},a_{32}\}=0,\quad \{a_{31},x_1\}=0,\nn\\ 
&[a_{31},x_2]=0,\quad \{a_{31},x_3\}=x_1,\quad \{a_{31},y_1\}=-y_3,\quad [a_{31},y_2]=0,\quad \{a_{31},y_3\}=0,\nn\\
&[a_{32},x_1]=0,\quad \{a_{32},x_2\}=0,\quad \{a_{32},x_3\}=x_2,\quad [a_{32},y_1]=0,\quad \{a_{32},y_2\}=-y_3,\nn\\ 
&\{a_{32},y_3\}=0,\quad \{x_1,x_2\}=2y_3,\quad \{x_1,x_3\}=-2y_2,\quad [x_1,y_1]=h_1+3h_2,\quad \{x_1,y_2\}=-3a_{21},\nn\\ 
&\{x_1,y_3\}=3a_{31},\quad \{x_2,x_3\}=-2y_1,\quad \{x_2,y_1\}=-3a_{12},\quad [x_2,y_2]=h_1,\quad \{x_2,y_3\}=3a_{32},\nn\\
&\{x_3,y_1\}=-3a_{13},\quad \{x_3,y_2\}=-3a_{23},\quad [x_3,y_3]=2h_1+3h_2,\quad \{y_1,y_2\}=2x_3,\quad \{y_1,y_3\}=-2x_2,\nn\\
&\{y_2,y_3\}=-2x_1.
\end{align}
So this yields a $\Z_2^2$-graded color Lie algebra compatible with the $\Z_2^2$-graded ordinary Lie algebra $G_2$, 
with sign factor given by
\[
\langle \alpha,\beta \rangle = \alpha_1\beta_2+\alpha_2\beta_1, \qquad (\alpha,\beta \in\Z_2\times\Z_2).
\]

\setcounter{equation}{0}
\section{Concluding remarks} 
\label{sec:E}%

In this paper we have studied the natural $\Z_2^3$-grading of the Lie algebra $G_2$.
For this grading, there is natural basis consisting of elements $A_\alpha^\zeta$, 
where $\alpha\in\Gamma^*=\Z_2^3\setminus\{000\}$ is the degree of the basis element,
and where $\zeta\in\alpha^\perp$. 
This basis is overcomplete, in the sense that $\sum_{\zeta\in\alpha^\perp} A_\alpha^\zeta=0$ for each $\alpha$. 
But on the other hand, the commutation relations in this basis follow a uniform pattern.
This basis of $G_2$ has a typical representation in the space of $8\times 8$-matrices ${\sf M}$.
Also ${\sf M}$ has a natural $\Z_2^3$-grading as a matrix algebra, and this grading is consistent with the grading of the $G_2$-basis.

This grading of $G_2$ allowed us to study $\Z_2^3$-graded color algebras of type $G_2$.
The possible sign factors for such colorings were determined in~\cite{Balbino2024}.
Here, we have presented (in all detail) three different $\Z_2^3$-graded color algebras of type $G_2$,
all three $\Z_2^3$-graded Lie algebras (and not $\Z_2^3$-graded Lie superalgebras) in the terminology of~\cite{Balbino2024}.

We found these three algebras by working through all possible sign changes in the matrix form of the graded basis elements of $G_2$,
and then looking for $\Z_2^3$-graded color algebras compatible with $G_2$ in the sense of~\eqref{compatible}.
In this regard, we cannot claim that we have a complete list of $\Z_2^3$-graded color algebras of type $G_2$.

We also show that a Cartan-Weyl basis of $G_2$, in which the Cartan basis elements are diagonal as $7\times 7$-matrices,
allows a $\Z_2^2$-grading. 
Contrary to the $\Z_2^3$-grading, the $\Z_2^2$-grading is compatible with the root structure of $G_2$.
Following the same technique as before, we could also construct a $\Z_2^2$-graded color algebra of type $G_2$, 
which is again a $\Z_2^2$-graded Lie algebra in the terminology of~\cite{Balbino2024}.

The results given in this paper are very explicit, both in giving matrix forms of basis elements, 
and in giving complete lists of commutators and/or anti-commutators. 
This seems a bit overdone.
On the other hand, we are dealing with new structures, and we think it can be useful to lay hands on such explicit results in order to have a better understanding.

The exceptional Lie algebra $G_2$ and the exceptional Lie superalgebra $G(3)$ have been used 
as symmetry algebra for ${\cal N}=7$ superconformal quantum mechanics~\cite{Toppan-G3}.
It would be worthwhile to study $\Z_2^n$-graded versions of this, in the sense of~\cite{Bruce2,AMD2020,Aizawa5,Quesne2021,Aizawa6}.
The current results for $G_2$ should be applicable for this.
Our work also opens the way to study gradings and colorings of $G(3)$, in particular gradings of type $\Z_2^n$.

\section*{Acknowledgments}

Both authors were supported by the Bulgarian National Science Fund, grant KP-06-N88/3.
The authors would also like to thank Francesco Toppan (Brazilian Center for Research in Physics)
for valuable suggestions.

\newpage
\setcounter{equation}{0}
\renewcommand{\theequation}{A.{\arabic{equation}}}
\section*{Appendix} 
\label{sec:F}%

\begin{center}
Table of brackets among the 14 basis elements of $G_2$, as given in~\eqref{g2-h}-\eqref{g2-x}.
\end{center}

\addtolength{\tabcolsep}{-2pt}  
\begin{center}
{\small
\begin{tabular}{c||c|c|c|c|c| c|c|c| c|c|c| c|c|c|}
 $[\cdot,\cdot]$    & $h_1$ & $h_2$ & $a_{12}$ & $a_{13}$ & $a_{23}$ & $a_{21}$ & $a_{31}$ & $a_{32}$ & $x_1$ & $x_2$ & $x_3$ & $y_1$ & $y_2$ & $y_3$ \\ \hline\hline
$h_1$ & $0$ & $0$ & $-3a_{12}$ & $0$ & $3a_{23}$ & $3a_{21}$ & $0$ & $-3a_{32}$ & $x_1$ & $-2x_2$ & $x_3$ & $-y_1$ & $2y_2$ & $-y_3$ \\ \hline
$h_2$ &  & $0$ & $2a_{12}$ & $a_{13}$ & $-a_{23}$ & $-2a_{21}$ & $-a_{31}$ & $a_{32}$ & $-x_1$ & $x_2$ & 0 & $y_1$ & $-y_2$ & 0 \\ \hline
$a_{12}$ &  &  &  0 & 0 & $a_{13}$ & $h_2$ & $-a_{32}$ & 0 & $-x_2$ & 0 & 0 & 0 & $y_1$ & 0 \\ \hline
$a_{13}$ &  &  &  & 0 & 0 & $-a_{23}$ & $h_1\!+\!2h_2$ & $a_{12}$ & $-x_3$ & 0 & 0 & 0 & 0 & $y_1$ \\ \hline
$a_{23}$ &  &  &  &  & 0 & 0 & $a_{21}$ & $h_1\!+\!h_2$ & 0 & $-x_3$ & 0 & 0 & 0 & $y_2$ \\ \hline
$a_{21}$ &  &  &  &  &  & 0 & 0 & $-a_{31}$ & 0 & $-x_1$ & 0 & $y_2$ & 0 & 0 \\ \hline
$a_{31}$ &  &  &  &  &  &  & 0 & 0 & 0 & 0 & $-x_1$ & $y_3$ & 0 & 0 \\ \hline
$a_{32}$ &  &  &  &  &  &  &  & 0 & 0 & 0 &$-x_2$ & 0 & $y_3$ & 0 \\ \hline
$x_1$    &  &  &  &  &  &  &  &  & 0 & $2y_3$ & $-2y_2$ & $h_1\!+\!3h_2$ & $3a_{21}$ & $3a_{31}$ \\ \hline
$x_2$    &  &  &  &  &  &  &  &  &  & 0 & $2y_1$ & $3a_{12}$ & $h_1$ & $3a_{32}$ \\ \hline
$x_3$    &  &  &  &  &  &  &  &  &  &  & 0 & $3a_{13}$ & $3a_{23}$ & $-2h_1\!-\!3h_2$ \\ \hline
$y_1$    &  &  &  &  &  &  &  &  &  &  &  & 0 & $2x_3$ & $-2x_2$ \\ \hline
$y_2$    &  &  &  &  &  &  &  &  &  &  &  &  & 0 & $2x_1$ \\ \hline
$y_3$    &  &  &  &  &  &  &  &  &  &  &  &  &  & 0 \\ \hline
\end{tabular}
}
\end{center}
\addtolength{\tabcolsep}{2pt}

\end{document}